\documentclass[useAMS,usenatbib]{mn2e}

\usepackage{times}

\newcommand{\iraf}  {{\sc iraf}}
\newcommand{\midas}  {{\sc midas}}

\usepackage{graphicx}
\usepackage{txfonts}
   \title[Optical Observations of PSR J0205+6449]
   {Optical Observations of PSR J0205+6449}

\author[A.\,Shearer and V.\,V.\,Neustroev]{A.\,Shearer\thanks{E-mail:
andy.shearer@nuigalway.ie; vitaly.neustroev@nuigalway.ie} and V.\,V.\,Neustroev\\
Centre for Astronomy, National University of Ireland, Galway, Newcastle Rd., Galway, Ireland\\
}

\begin{document}

\date{Accepted 2008 July 3.  Received 2008 July 3; in original form 2008 June 18}

\pagerange{\pageref{firstpage}--\pageref{lastpage}} \pubyear{2008}

\maketitle

\label{firstpage}

\begin{abstract}

PSR J0205+6449 is a X-ray and radio pulsar in supernova remnant 3C 58.  We report on observations of the central region of 3C 58 using the 4.2-m William Herschel Telescope with the intention of identifying the optical counterpart of PSR J0205+6449 and characterising its pulsar wind nebula.
 Around the pulsar position we identified extended emission with a magnitude of $B = 23 \fm 97 \pm 0.10$, $V = 22 \fm 95 \pm 0.05$ and $R = 22 \fm 15 \pm 0.03$ consistent with a pulsar wind nebula. From the R-band image we identified three knots with $m_R$ = $24 \fm 08 \pm 0.07$ (o1), $24 \fm 15 \pm 0.07$ (o2) and $24 \fm 24 \pm 0.08$ (o3).
We confirm the presence of an optical pulsar wind nebula around PSR J0205+6449 and give an upper limit of $m_R \approx$ 24 for the optical magnitude of the pulsar. Furthermore we make the tentative suggestion that our object o1, with an $m_R \approx$ 24.08 is the optical counterpart. If confirmed the pulsar would have an $L_R/L_x\approx 0.004$ and an optical efficiency of about 5\% of the Crab pulsar. Such a low efficiency is more consistent with the characteristic age of the pulsar rather than that of SN 1181.
\end{abstract}


%

\section{Introduction}

PSR J0205+6449 in supernova remnant (SNR) 3C58  is a recently discovered X-ray and radio pulsar \citep{ca02,mu02} with a period of 65 ms.  3C 58  was thought to be young due to a possible  association with SN 1181 \citep{vb78}. Consequently it should share many of the characteristics of the Crab nebula, including the presence of a young pulsar. However an embedded pulsar  defied detection for over twenty years. PSR J0205+6449 has a considerably weaker flux than the Crab pulsar. Its X-ray emission is 1000 times lower than the Crab and its radio emission 120 times lower. Some of this discrepancy can be attributed to the mounting evidence that the characteristic age of the pulsar  
($P / 2\dot{P} $ $\approx$ 5400 years) is near to its true age and hence it is not associated with SN 1181 \citep{go07}.  VLA observations at 1.4GHz show a nebula expansion rate of $0.014\% \pm 0.003$ yr$^{-1}$ \citep{bi06},  inconsistent with SN 1181 unless substantial deceleration of the remnant has occurred.  A comprehensive optical study of 3C 58 showed no evidence of an optical pulsar at m$_R \approx $22.5 \citep{Fesen08}. Furthermore based upon the relatively low proper motion of knots within 3C 58 these authors also cast doubt on the association between 3C 58 and SN 1181. Deeper optical observations of 3C 58 show evidence of an optical nebulosity at the same location as the X-ray counterpart to PSR J0205+6449 \citep{Shibanov08}. These authors interpret this nebulosity as a pulsar wind nebula (PWN). 

PSR J0205+6449  has the third highest spin down energy flux [$\dot{E}/d^2 \approx 2.6~10^{36}$ ergs/s/kpc$^2$] after the Crab and Vela pulsars. The X-ray determined hydrogen column density [$N_H\approx 4.1~10^{21} cm^{-2}$], \citep{go07} and optical extinction, E(B-V)  $\sim$ 0.68 \citep{Fesen08}, are similar to the Crab pulsar.  From H{\small I} observations a kinematic distance of 3C 58  has been established at  3.2 kpc  \citep{ro93}, consistent with its Dispersion Measure \citep{ca02}. These combine to make  PSR J0205+6449 a likely candidate for optical emission studies. If we assume that the optical luminosity scales with light cylinder magnetic field,  $L_{opt} \propto B_{lc}^{1.6}$ \citep{sg02}, then we estimate that the pulsar should have a visual magnitude in the range 23--25, depending on interstellar absorption and effects of beaming geometry \citep{sh08b}. This paper reports on a William Herschel telescope (WHT) service time observation of  PSR J0205+6449 to look for evidence of an optical pulsar and its possible PWN. 

\section{Observations and data reduction}
\subsection{Observations}
Photometric observations of the field surrounding 3C~58 were performed on the night of September 10, 2007
using the 4.2-m WHT at the Isaac Newton Group of Telescopes, La Palma.
The observations were taken in Service mode with the Auxiliary Port Imaging Camera (AUX), using
the Harris set of the broadband BVR filters. It is thought that when the filter response curves are convolved
with a typical CCD response, these glass filters provide a closer match to the standard Johnson $B$ and $V$ and
the Kron-Cousins $R$ bandpass. AUX, with its 1024$\times$1024 TEK CCD, has an unvignetted,
circular field diameter of 1.8 arcmin with the pixel size on the sky of 0\farcs108 $\times$ 0\farcs108. The integration time was 1000 s in the $B$ and $V$ bands each, and 1400 s in the $R$ band. The Landolt standard star 94-242 was observed
immediately after the primary target at a similar airmass. The weather conditions were very good and stable
during the observations. As a result, the $FWHM$ of a stellar object in the images is less than 0\farcs75 in
all the bands. Table \ref{ObsTab} provides a journal of the observations. The data were reduced using standard techniques within \iraf\ and \midas. The frames were debiased and flat fielded
using sky flats from the same night.

\begin{table}
\caption[]{Summary of Observations}
\label{ObsTab}
\begin{flushleft}
\begin{tabular}{cccccc}
\hline\noalign{\smallskip}
 Date          &  UTC    & Filter & Duration &  sec$z$  &  Seeing\\
               & \null   &  \null & (s)      &          & ($\arcsec$)       \\
\noalign{\smallskip}
\hline\noalign{\smallskip}
2007-Sep-11   &   02:34:27 &  B & 1000   & $1.31$ & $0.75$ \\
2007-Sep-11    &  02:54:27 &  V & 1000   & $1.25$ & $0.75$ \\
2007-Sep-11    &  03:13:50 &  R & 1400   & $1.25$ & $0.70$ \\
\noalign{\smallskip}
\hline
\end{tabular}
\end{flushleft}
\end{table}
\begin{table}
\begin{center}
\caption[]{The secondary standard stars used for photometric referencing.}
\label{CompStarsTab}
\begin{tabular}{cccc}
\noalign{\smallskip}
\hline
\noalign{\smallskip}
Star&   $B$  &  $V$    &  $R$  \\
    &   mag  &  mag    &  mag  \\
\noalign{\smallskip}
\hline\noalign{\smallskip}
1   & 19.73  & 18.77  & 18.15  \\
2   & 20.60  & 19.36  & 18.58  \\
3   & 22.25  & 20.79  & 19.86  \\
4   & 21.29  & 19.74  & 18.76  \\
5   & 21.96  & 20.73  & 19.98  \\
\noalign{\smallskip}
\hline
\end{tabular}
\end{center}
The formal, statistical errors are less than 0\fm01 for all the stars with the exclusion of the stars 3 and 5 in the $B$ band where they have errors of $\sim$0\fm02. 

\end{table}

\subsection{Astrometry}
Our astrometry was made using positions
of ten reference stars from the USNO-B1.0 catalogue\footnote{The stars from the USNO-B1.0 catalogue used for the
astrometry: 1548-0060184, 1548-0060191, 1548-0060202, 1548-0060210,
1548-0060227, 1548-0060230, 1548-0060232, 1548-0060256, 1548-0060258, 1548-0060300, 1548-0060315.}.
We used the \iraf\ tasks \textsc{ccmap/cctran} for the astrometric transformation of the images.
Formal rms errors of the astrometric fit for the RA and DEC were 0\farcs066 and 0\farcs059 for the R,
0\farcs081 and 0\farcs035 for the V, and 0\farcs079 and 0\farcs041 for the B bands respectively, that is better
than the pixel size of the images.

\begin{figure*}
\centering
\includegraphics[width=12cm]{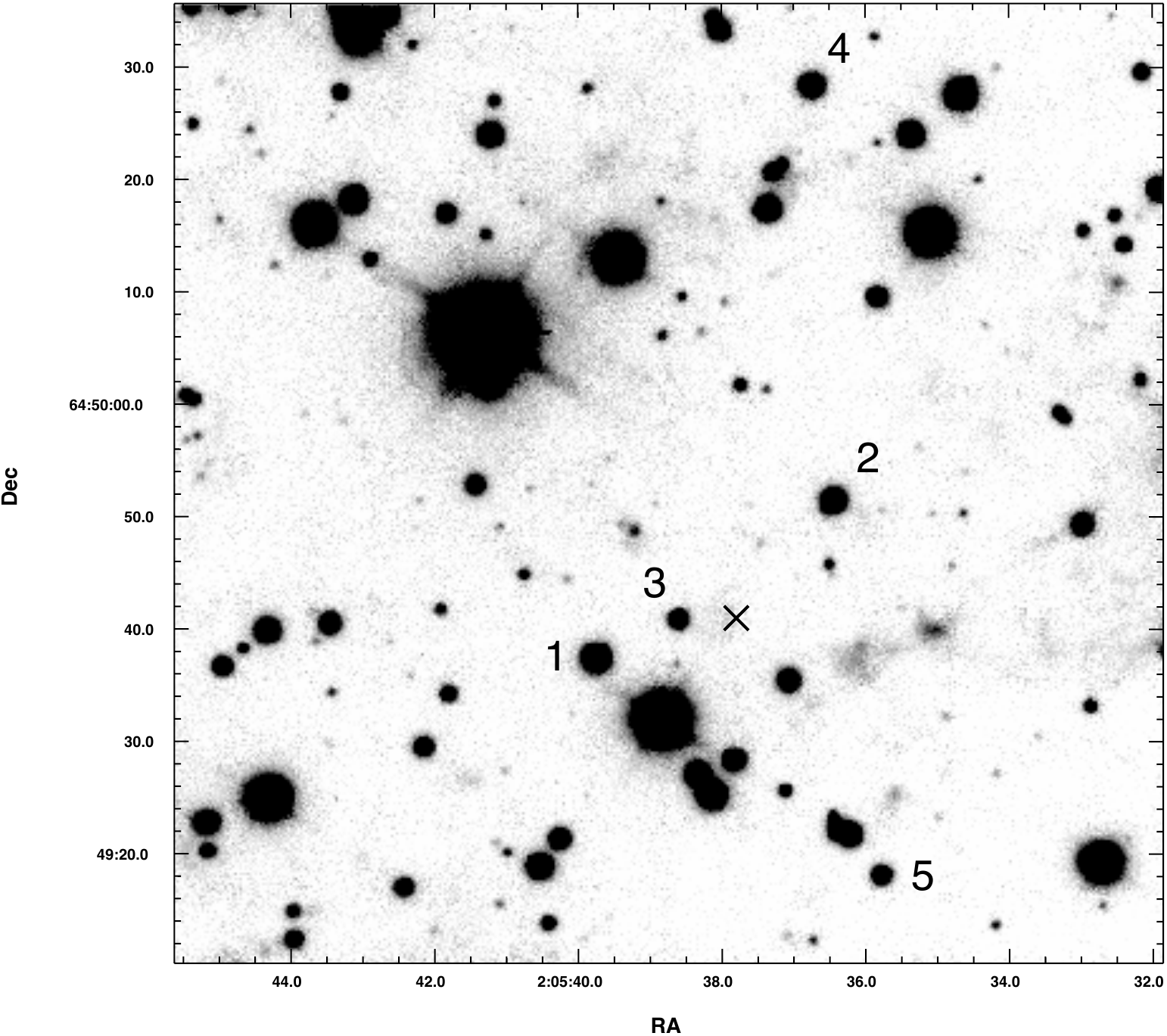}
 \caption{R-band image of the central region of 3C~58. The Chandra position of the pulsar PSR J0205+6449
 \citep{mu02} is marked by $\times$. The numbers mark the secondary photometric standard stars
 from Table~2.}
 \label{CompStarsFig}
\end{figure*}

\subsection{Photometry}
First of all, the magnitudes of five relatively bright stars visible in the target frame were derived accurately
(Figure~\ref{CompStarsFig}).  This photometric calibration was carried using the Landolt standard star 94-242.
Our data do not allow us to determine the atmospheric extintion coefficient. Instead of this we use the average coefficients $k_B = 0.21$, $k_V = 0.13$ and $k_R = 0.09$, provided for the Roque de Los Muchachos Observatory,
La Palma \citep{Kidger}.
The signal-to-noise ratios $S/N$ and the magnitude uncertainties $\Delta m$ were calculated as
\begin{equation}
\frac{S}{N} =
\frac{f_{ap}}{\sqrt{f_{ap}/g+n_{ap}\sigma_{bg}^2(1+1/n_{bg})}}
\end{equation}
\begin{equation}
\Delta m = 1.0856\ \Bigl(\frac{S}{N}\Bigr)^{-1},
\end{equation}
where $f_{ap}$ is the source flux in counts for a given aperture,
$g$ is the gain,
$n_{ap}$ the number of pixels in the source aperture,
$n_{bg}$  the number of pixels in area used for the background measurement.
and $\sigma_{bg}$ is the standard deviation of the background in counts \citep{Newberry}. The stars chosen as secondary standards are marked by numbers in Figure~\ref{CompStarsFig},
and their magnitudes with errors are listed in Table~\ref{CompStarsTab}. The resulting zero-points for the pulsar frames are $B=25 \fm 97$,
$V=26 \fm 06$ and $R=26 \fm 20$. The formal, statistical zero-point errors are less than 0\fm01.
However, since only an average curve for atmospheric extinction was used and only one standard star was observed and this only once, the real errors can be as large as few $\times$ 0.01 magnitude.

\begin{figure*}
\centering
\includegraphics[width=8.5cm]{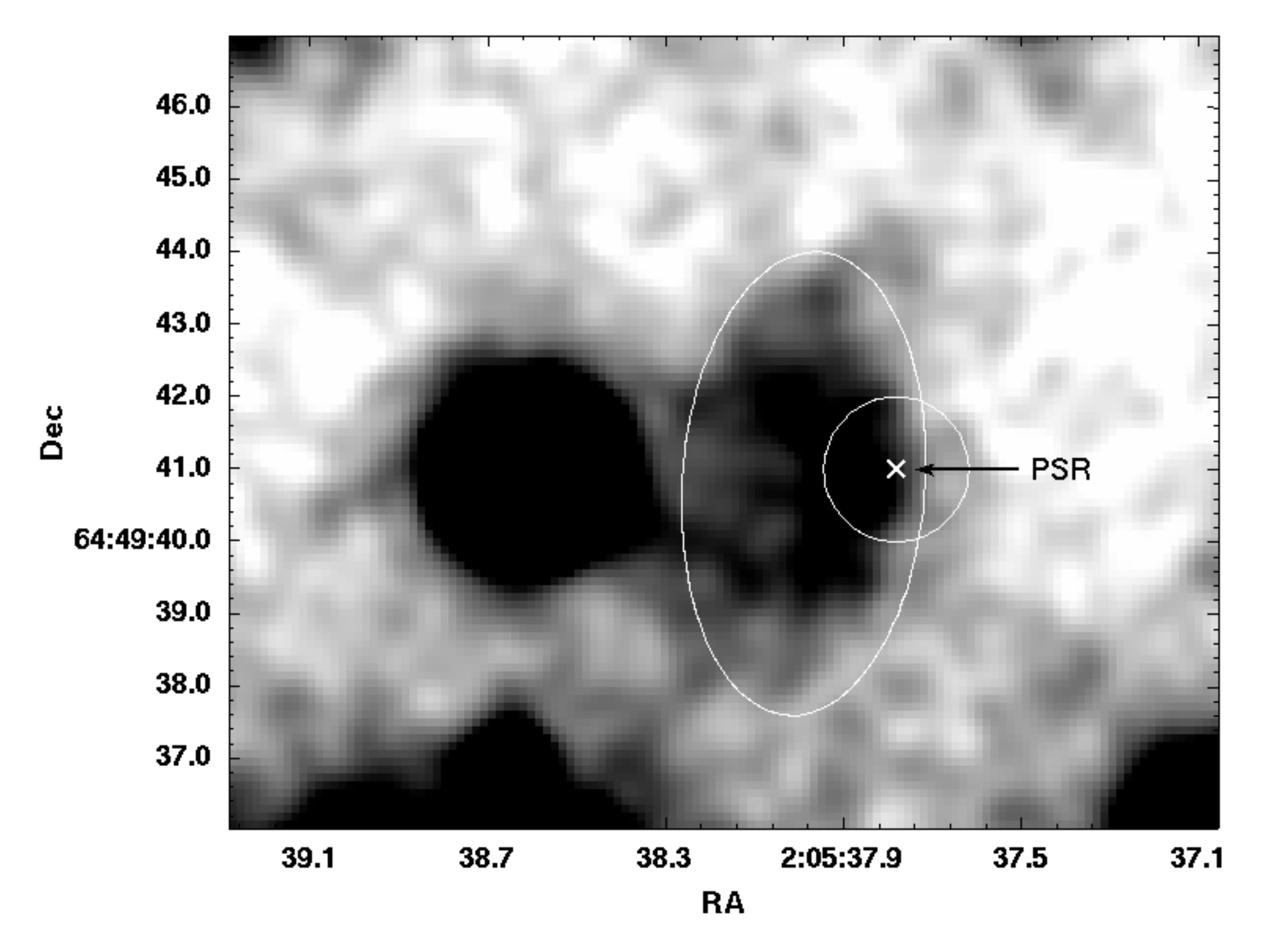}
\includegraphics[width=8.5cm]{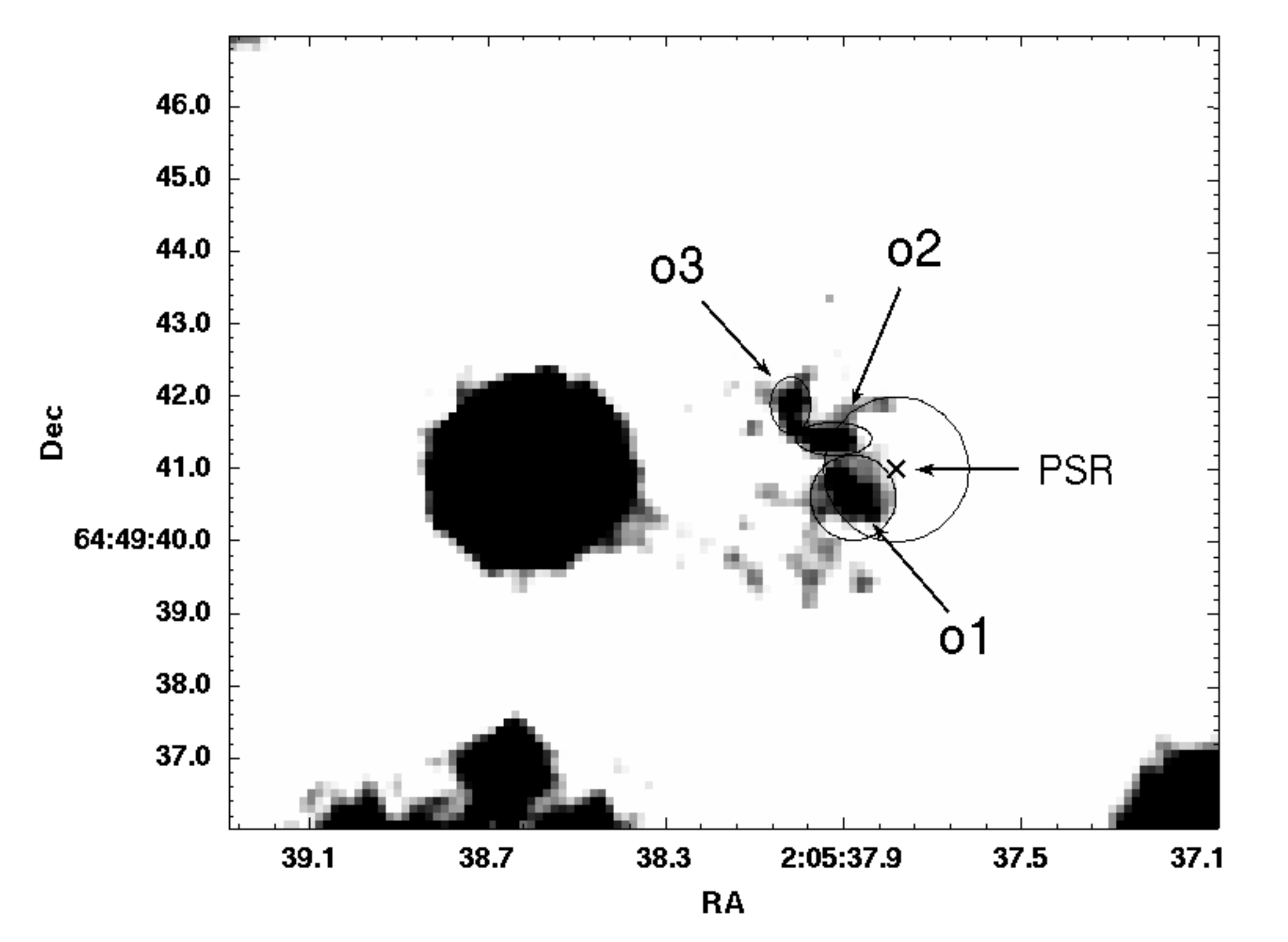}
  \caption{Fragment of the R image shown with different contrast to emphasize different components of the nebula.
  The $\times$ symbol marks the Chandra position of the pulsar PSR J0205+6449. The typical pointing uncertainty,
  $\lesssim 1 \arcsec$ is shown by the circle. In the left panel we show the elliptical aperture used for photometry of the nebula. In the right panel we mark three compact objects inside the nebula.}
\label{psrFig}
\end{figure*}

\begin{figure*}
\centering
\includegraphics[width=8.8cm]{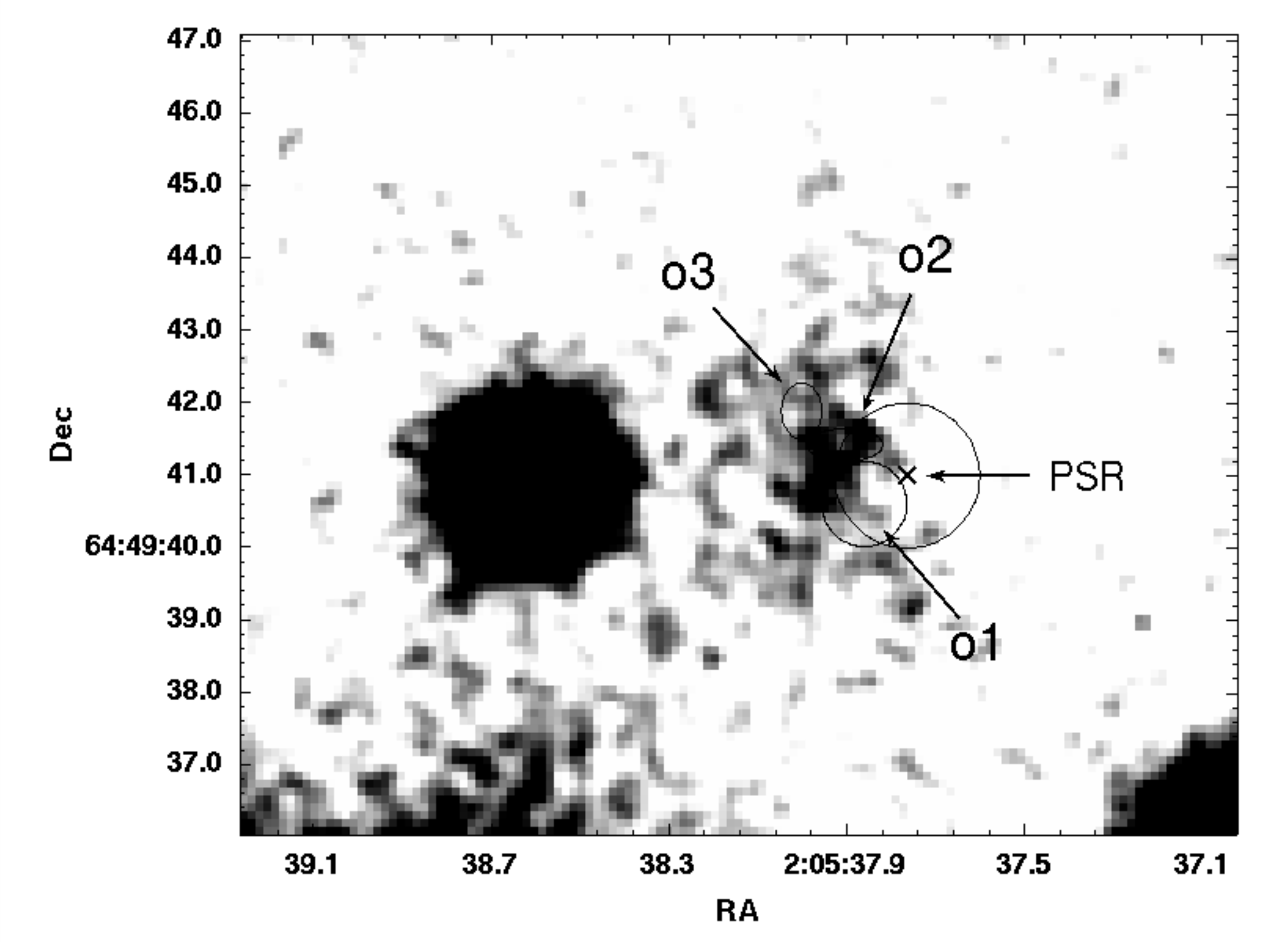}
\includegraphics[width=8.8cm]{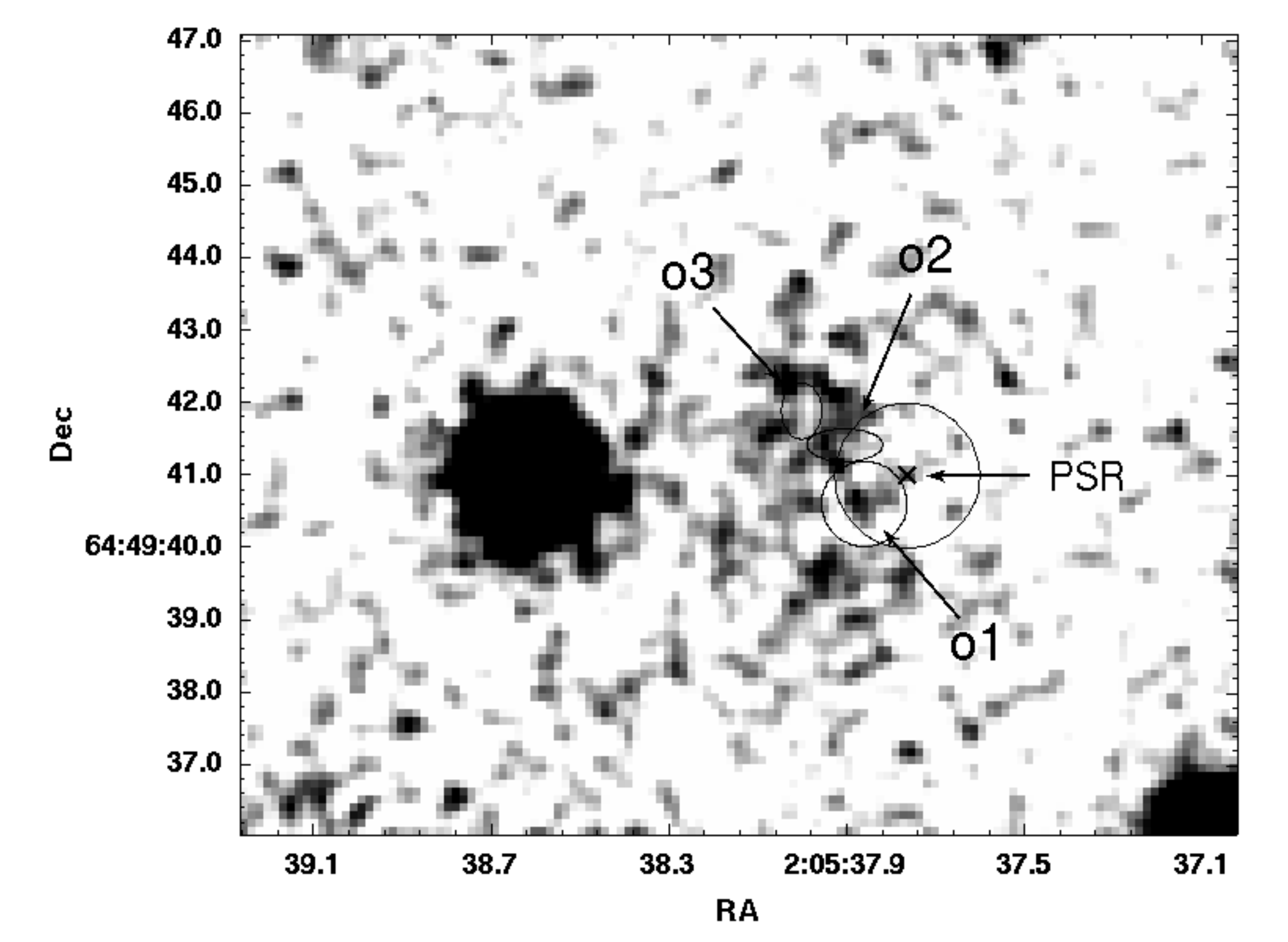}
  \caption{High contrast V (left) and B (right)  images of same region shown in Figure \ref{psrFig} }
\label{psrFig2}
\end{figure*}

\section{Morphology of the 3C~58 field}
Recent observations of 3C~58 \citep{Shibanov08} have shown evidence of an optical nebulosity
at the same location as the X-ray counterpart of PSR J0205+6449. They described it as a faint extended, elliptical
structure with no resolved point-like object at its centre. Our observations confirm the presence of this
nebulosity. However, its brightest area has a non-elliptical shape with an arc-like filamentary structure.
Figure~\ref{psrFig} (left panel) shows the R band image of this region. In fact, such a
nebula structure is also clearly seen in the V image of 3C~58 presented by \citet{Shibanov08}
in their Figure~3. For consistency with Shibanov et al., we have performed
photometry of this area using the similar elliptical apertures (Table 1 in their paper). The aperture that formally 
encapsulates $\gtrsim$ 86\% of the total nebula flux, is shown in Figure~\ref{psrFig} (left panel) as a white ellipse.
The measured integral magnitudes of the nebula are $B = 23 \fm 97 \pm 0.10$, $V = 22 \fm 95 \pm 0.05$ and
$R = 22 \fm 15 \pm 0.03$. The measured B and V magnitudes agree within errors with those of \citet{Shibanov08}.
although our values are a bit brighter. The latter can be explained by non-perfect reduction of the standard stars,
as well as by the variable background. We note however that in the R band the nebula is brighter than the
upper limit for the pulsar magnitude taken from \citet{Fesen08}.

Furthermore, the R image shows this nebula with some structure significantly above the noise.
In the right panel of Figure~\ref{psrFig} we show the same area of the R image, but with different contrast to 
emphasise different components of the nebula. There are three compact objects in the middle of this nebula, marked 
in Figure~\ref{psrFig} as o1, o2 and o3 (the last two could be one arc-like object).
We have measured the magnitudes of these objects, using a 5 pixel radius  aperture corresponding to a sky radius of approximately 0\farcs54. We used the same background as determined previously for the photometry of the nebula.

The measured magnitudes are $24 \fm 08 \pm 0.07$ (o1), $24 \fm 15 \pm 0.07$ (o2) and $24 \fm 24 \pm 0.08$ (o3).
We have also measured the flux from the brightest part of the nebula outside the objects, using the same aperture,
and obtained the value $\approx 24 \fm 55 \pm 0.10$, giving us confidence in the reality of the compact objects in the 
nebula. Unfortunately, the B and V images are too noisy to confirm their existence in these bands although upper limits can be given. The high contrast B and V and images are shown in Figure~\ref{psrFig2}. Based upon \citet{car89} we expect E(V-R) and E(B-R) to be $\sim$ 0.5 and 1.2 respectively, and this extinction probably accounts for the non-detection in the B and V bands.   The results of our
photometry, with the measured coordinates of the compact objects inside of the nebula, are presented in Table \ref{Fluxes}.

In conclusion we note that the object o1 is well inside the typical HRC pointing uncertainty, $\lesssim 1 \arcsec$, marked in Figure~\ref{psrFig} as a circle around the pulsar position. We propose that this is the best candidate for the optical
counterpart of the pulsar.

\begin{table}
\caption[]{Fluxes and positions of filamentary  structures.} 
\begin{center}
\label{Fluxes}
\begin{tabular}{cccccc}
\noalign{\smallskip}
\hline
\noalign{\smallskip}
Knot&  RA & Dec &  $B$  &  $V$    &  $R$  \\
      & (h m s) & (\degr \arcmin \arcsec) &   mag  &  mag    &  mag  \\
\noalign{\smallskip}
\hline\noalign{\smallskip}
o1 & 02 05 37.90 & +64 49 40.6 & $>$26.1  & $>$24.7  & 24.08 \\
o2 & 02 05 37.93 & +64 49 41.4 & $>$25.6  & $>$24.3  & 24.15  \\
o3  & 02 05 38.02 & +64 49 41.9  & $>$25.5  & $>$24.7  & 24.24  \\
\noalign{\smallskip}
\hline
\end{tabular}

\end{center}

\end{table}

\section{Conclusions}

From these observations we confirm the presence of an optical PWN around PSR J0205+6449 and identify possible structure within the nebula.
Although from these observations we can not definitively identify the optical counterpart of PSR J0205+6449 we can give a deep upper limit and make a tentative suggestion that our object o1 is  the counterpart. If it is shown that o1 is the counterpart then we can estimate, using \citet{fg94}, the flux from the pulsar to be 0.71 $\mu$Jy in the range $\lambda\lambda$ 6000-7000 \AA, corresponding to a R-band luminosity of $7.3~10^{30}$ ergs sec$^{-1}$ based upon an E(B-V)=0.68 giving A$_R \approx$ 1.8. This luminosity gives an efficiency  $L_R/\dot{E}$ of $\approx 3~10^{-7}$ and a $L_R/L_x\approx 0.004$. The optical efficiency is about 5\% of the Crab pulsar's optical efficiency and comparable to the efficiency of older optical pulsars such as Vela and PSR B0656+14 \citep{sg02}. The optical -- X-Ray luminosity ratio is relatively high at 0.004, but this is more of a reflection of the low X-ray luminosity. From these tentative observations we can conclude that the optical emission, if confirmed, is consistent with a pulsar with an age similar to its spin down age rather than that of a very young pulsar. Final optical identification will have to wait for observations sensitive to flux variations on millisecond time scales using  instruments such as Optima \citep{st01} or GASP \citep{co08}.

\section*{Acknowledgements}
We thank the ING Group of Telescopes for providing these service time observations and in particular the support astronomer, Dr. Chris Benn, who took the observations. VN acknowledges the support of both Science Foundation Ireland (grant number 02/IN.1/I208 WebCom-G) and the Higher Education Authority, through its CosmoGrid programme. We also thank the anonymous referee for comments that improved an earlier draft of this manuscript.

\label{lastpage}
\end{document}